# Recent Advances in Thermoelectric Performance of Half-Heusler Compounds

S. Joseph Poon

Department of Physics, University of Virginia, Charlottesville, Virginia 22904-4714, VA, USA; sjp9x@virginia.edu; Tel.: +1-434-924-6792

**Abstract:** Half-Heusler phases (space group $F\bar{4}3m$, C1$_b$) have recently captured much attention as promising thermoelectric materials for heat-to-electric power conversion in the mid-to-high temperature range. The most studied ones are the RNiSn-type half-Heusler compounds, where R represents refractory metals Hf, Zr, and Ti. These compounds have shown a high-power factor and high-power density, as well as good material stability and scalability. Due to their high thermal conductivity, however, the dimensionless figure of merit ($zT$) of these materials has stagnated near 1 for a long time. Since 2013, the verifiable $zT$ of half-Heusler compounds has risen from 1 to near 1.5 for both n- and p-type compounds in the temperature range of 500–900 °C. In this brief review, we summarize recent advances as well as approaches in achieving the high $zT$ reported. In particular, we discuss the less-exploited strain-relief effect and dopant resonant state effect studied by the author and his collaborators in more detail. Finally, we point out directions for further development.

**Keywords:** half-Heusler compounds; figure of merit; power density; lattice disorder; dopant resonant states

## 1. Introduction

Thermoelectric energy conversion is an important component of renewable-energy technology. Among many thermoelectric (TE) materials, half-Heusler compounds (space group $F\bar{4}3m$, C1$_b$), principally with the composition XYZ, where X and Y are transition or rare earth elements and Z is a main group element, are the most studied ones. Half-Heusler (HH) compounds possess several promising features, namely, a high Seebeck coefficient (also known as thermopower), moderate electrical resistivity, and good thermal stability [1–4]. As for scale-up, HH compounds can be produced in large quantities [5]. In particular, the RNiSn-type HH compounds, where R represents refractory metals Hf, Zr, and Ti, are the most studied to date. The dimensionless figure of merit, $zT$, for TE materials is defined as $zT = (S^2\sigma/k)T$, where $\sigma$ is the electrical conductivity, $S$ is the Seebeck coefficient, and $k$ is thermal conductivity. $zT$ can also be written as $zT = (PF^*T)/k$ where $PF = S\sigma^2$ is the power factor, "*" denotes and hereafter the $PFx^*T$ term, which has the same unit as $k$, is the power factor temperature product. Until 2013, the highest $zT$ of HH compounds was near 1 in the temperature range of 500–1000 K [6–8]. Meanwhile, the report of $zT$ near 1.5 at 600 K in the HH compound, Ti$_{0.5}$Zr$_{0.25}$Hf$_{0.25}$NiSn$_{0.998}$Sb$_{0.002}$, by Sakurada and N. Shutoh [9] more than a decade ago has not escaped notice; however, no independent verification was reported.

There are two basic approaches to enhancing the $zT$ of half-Heusler compounds: (1) lowering thermal conductivity, and (2) raising the power factor. These two TE properties are inter-related and it is nontrivial to simultaneously improve them. Nevertheless, by employing different approaches, several groups have been successful in overcoming the barrier to high $zT$ in HH compounds. In this brief review, we first summarize the notable results in half-Heusler compounds since 2013, namely, *zT* > 1, and the origin of high thermoelectric performance. Next, we provide a more detailed discussion on two less studied effects, namely, the strain-relief effect and dopant resonant state effect for improving the TE properties of HH compounds. Finally, we conclude with a summary and outlook.

## 2. Results and Discussion

### 2.1. High Dimensionless Figures of Merit (zT) in Half-Heusler Compounds

The highest *zT* (>1) reported for several n-type and p-type half-Heusler compounds are listed in Table 1. The origins of high *zT* in these HH compounds were attributed to various enabling factors, namely, structural order, microstructure including nanostructure, mass fluctuations, dopant-related resonant states, high band degeneracy, and heavy hole band. Each of these factors usually influences more than one specific TE property. For example, a nanostructure apparently influences both thermal conductivity and the Seebeck coefficient, while mass fluctuations due to alloying can influence both thermal conductivity and band structure. Clearly such complexity requires new thinking beyond the "phonon glass electron crystal" paradigm. As shown in Table 1, recent results indicated that researchers have been successful in optimizing *zT* in half-Heusler compounds. However, further fundamental studies are needed in order to advance to the next level.

**Table 1.** Half-Heusler compounds with their highest *zT* (>1) obtained to date. Included in the table are *zT* at the indicated temperatures and the enabling factors and effects. Enabling factors may already pre-exist in some compounds but in a less prominent way. Data are from the cited references.

| Compounds | *zT* (K) | Enabling factors | Effects | References |
|---|---|---|---|---|
| **N Type HH Compounds** | | | | |
| $Ti_{0.5}Zr_{0.25}Hf_{0.25}NiSn$ | 1.2 (830 K) | Phase separation | Lower *k* | [10] |
| $Hf_{0.6}Zr_{0.4}NiSn_{0.995}Sb_{0.005}$ | 1.2 (860 K) | Strain reduction | Higher *ρ* and *S*, lower *k* | [11] |
| $Hf_{0.65}Zr_{0.25}Ti_{0.15}NiSn_{0.995}Sb_{0.005}$ | 1.3 (830 K) | Embedded nano-oxide | Higher *ρ* and *S*, lower *k* | [12] |
| $Hf_{0.594}Zr_{0.396}V_{0.01}NiSn_{0.995}Sb_{0.005}$ | 1.3 (900 K) | Dopant resonant states | Higher *ρ* and *S*, lower *k* | [13] |
| $Ti_{0.5}Zr_{0.5}NiSn_{0.98}Sb_{0.02}$ $Ti_{0.5}Zr_{0.25}Hf_{0.25}NiSn_{0.98}Sb_{0.02}$ | ~1.2 (820 K) | Presence of nanograin | Higher *ρ* and *S*, lower *k* | [14] |
| $Ti_{0.5}Zr_{0.25}Hf_{0.25}NiSn$ | 1.5 (820 K) | Similar to Reference [14] | Higher *S*, lower *ρ* and *k* | [15] |
| **P-Type HH Compounds** | | | | |
| $FeNb_{0.88}Hf_{0.12}Sb$ $FeNb_{0.86}Hf_{0.14}Sb$ | 1.5 (1200 K) | Heavy hole band, high dopant content, heavy atomic mass | Lower *ρ*, low *k* Retain high *S* | [16] |
| $ZrCoBi_{0.65}Sb_{0.15}Sn_{0.20}$ | ~1.4 (970 K) | High band degeneracy, high dopant content, heavy atomic mass | Lower *ρ*, low *k* Retain high *S* | [17] |

*zT* versus *T* plots for state-of-the-art thermoelectric materials have been shown in several review articles, and are not reproduced here [6–8]. To date, half-Heusler compounds have clearly surpassed nano-SiGe in the temperature range above 800 K. HH compounds are now considered one of the few practical high *zT* materials, along with skutterudite and chalcogenide compounds. The competitiveness of half-Heusler compounds among the mentioned material classes, coupled with their excellent scalability, will no doubt continue to generate a plethora of research efforts due to the need for practical TE materials.

*2.2. Power-Conversion Efficiency of TE Modules Based on HH Compounds*

Earlier work on an n–p couple module built with $zT < 1$ n and p-type half-Heusler compounds reported a power-conversion efficiency (η) of near 9% [18] and power density ~8.9 W/cm² for a single-couple module, and ~3.5 W/cm² for a 49-couple module (unpublished results). η is near the ideal value of 9.2%, calculated using the expressions in Reference [12], as shown in Figure 1. The measured conversion efficiency is comparable to the state-of-the-art thermoelectric generators (TEGs), despite $zT < 1$. The observation of near-ideal η in modules based on $zT < 1$ HH compounds is certainly encouraging. More recently, a three-stage cascade TEG with a conversion efficiency of 20% was reported [19]. TEGs based on the high $zT > 1$ HH compounds discussed would have ideal conversion efficiency η near 13% [12,17]. The power-conversion efficiency and power density for an n-p couple and unileg devices built from HH compounds are summarized in Table 2. The potential for a large power output exists. However, the current conversion efficiencies of TEGs built on high $zT$ HH compounds still fall significantly below the ideal values. Such shortfall underlines the need for addressing outstanding material science and engineering issues in order to make the transition from high $zT$ materials to high η modules.

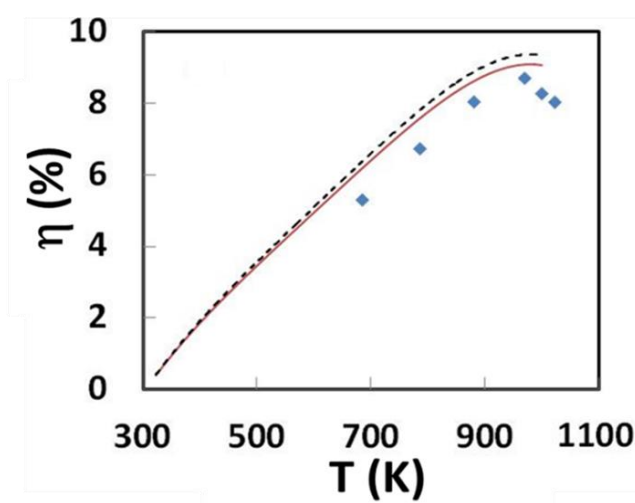

**Figure 1.** Thermoelectric (TE) conversion efficiency for n-p couple built using $Hf_{0.6}Zr_{0.4}NiSn_{0.995}Sb_{0.005}$ (n-type) and $Hf_{0.3}Zr_{0.7}CoSn_{0.3}Sb_{0.7}$/nano-$ZrO_2$ (p-type). Plots: measurement (cyan rhombus), calculation using temperature-averaged $zT$ (dashed line), calculation using temperature-dependent parameters in Reference [12] (solid line).

**Table 2.** Thermoelectric conversion efficiency and power density of n-p couple and unileg devices. $zT$ values of the n and p legs are also included. $\Delta T$ is the temperature difference between the hot and cold ends. Data are extracted from the references. Unpublished results are noted.

| $zT$ n-p Couple and Unileg Devices | $\Delta T$ (°C) | η (%) Measured/Ideal | Power Density (W/cm²) | References |
|---|---|---|---|---|
| 1 (n), 0.8 (p) | 700 | 8.7/9.2 | 8.9<br>3.5 (49-couple) | [18] (also unpublished) |
| ~1 (n), ~1.2 (p) | 650 | ~6/11 | 9.5 | [16] |
| 0.8 (n), 0.47 (p) | 527 | - | 3.2 (7-couple) | [5] |
| 0.45 (n) | 560 | - | 0.28 | [20] |
| 1.4 (p) | 500 | ~9/12 | ~9 | [17] |

*2.3. Mitigation of Lattice Strain to Enhance zT*

In this and the following sections, we highlight two methods employed by the author and his collaborators to improve *zT*. Specifically, the methods exploit lattice-strain relief and dopant resonant states. The figures shown herein are either new or adapted from recent publications. The unit cell of a half-Heusler phase consists of three occupied face-centered cubic (fcc) sublattices and an unoccupied fcc sublattice, as shown in Figure 2 below.

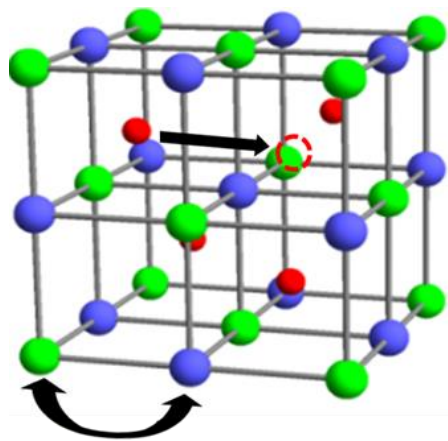

**Figure 2.** Unit cell of Hf (●)Sn (●)Ni (●). Antisite disorder includes occupancy of supposedly vacant fcc site by Ni and atom swap between Sn and Hf.

Synthesis and characterization of the samples studied have been described in References [11–13]. Worth special mentioning is that, since dopant concentrations are usually lower than 1 at.%, it is important to confirm the accuracy of the compositions of the doped samples. In the case of (Hf,Zr)NiSn doped with less than 1 at.% Sb or V. The alloy ingots were arc-melted using proper amounts of elemental Hf, Zr, Ni, and Sn, and premelted $Sn_{90}Sb_{10}$ or $Hf_{90}V_{10}$. Latter precursor alloys that contained much higher Sb and V contents were used in order to minimize compositional uncertainty. Energy-dispersive spectroscopy (EDS) was then performed to obtain the elemental distributions in the final ingot. The compositional homogeneity checked by EDS map scan over a $9 \times 10^4$ μm$^2$ area did not detect any compositional variation across a few μm$^2$ [11].

Antisite disorder is inevitable due to the close structural relationship between the half-Heusler phase and full-Heusler phase [21,22]. The likely antisite disorder involves the transfer of an Ni atom from an occupied fcc sublattice site to one of the vacant fcc sublattice sites, as shown in Figure 2. However, atom swaps between two occupied sites, e.g. between Hf and Sn sites, are also possible. The study of the effect of antisite disorder on TE properties has recently gained much attention [23–29]. The electronic structure that determines effective mass and mobility, as well as phonon scattering that gives rise to lattice thermal conductivity, are sensitive to the type of structural disorder in half-Heusler compounds [29]. Antisite disorder inevitably causes lattice strain. The level of strain can be determined from the peak widths in X-ray diffraction. Our study showed that the strain in HH compounds could be reduced if the sintering temperature in spark plasma sintering (SPS) was increased to near the melting temperature [11]. Normally, the compounds were sintered at lower temperatures. The X-ray pattern for $Hf_{0.6}Zr_{0.4}NiSn_{0.995}Sb_{0.005}$ is shown in Figure 3a. Typical peak widths of the (220) reflection obtained on samples prepared under different conditions are shown in the inset. Figure 3b illustrates the standard equation used in analyzing lattice strain ε. The strain was found to precipitously decrease with increasing processing temperature. As shown in Figure 3c, the level of strain in the samples decreases with increasing synthesis temperature, from samples annealed at 700 °C to SPS samples compacted near the melting point (~1450 °C), by as much as 30%.

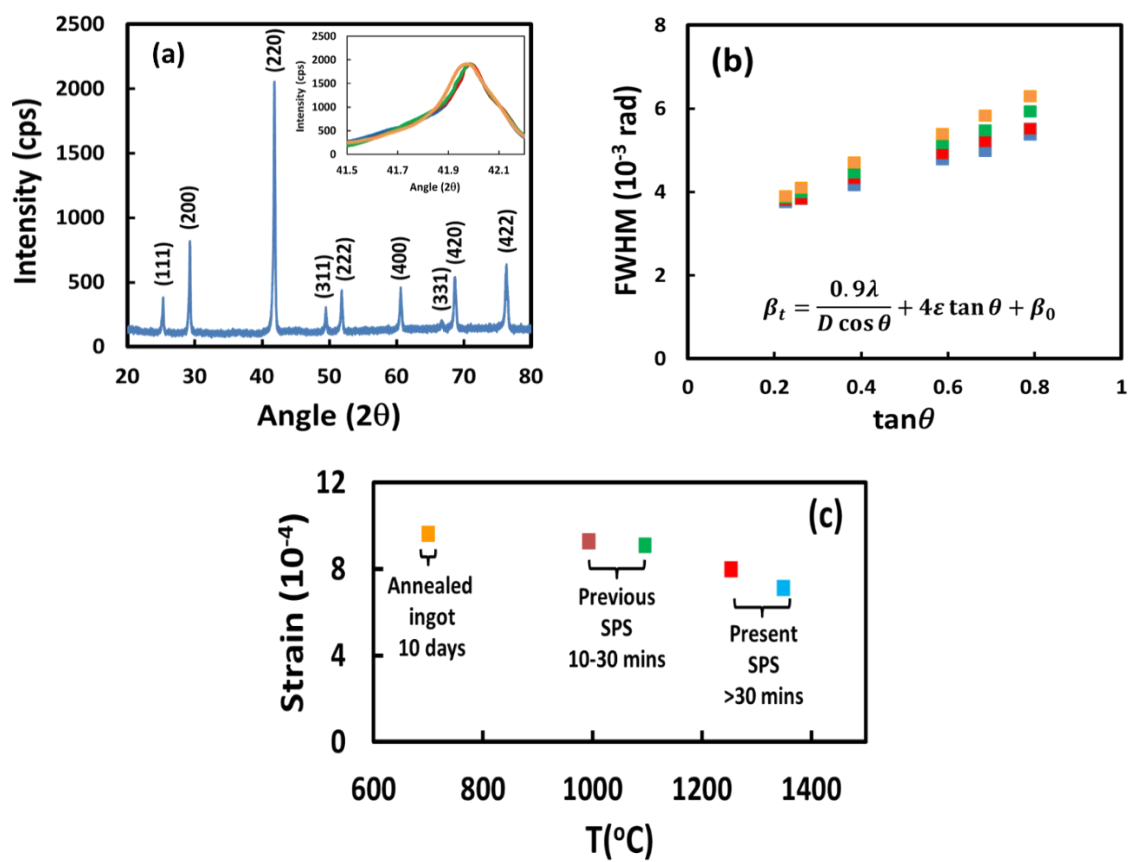

**Figure 3.** (**a**) X-ray pattern for Hf$_{0.6}$Zr$_{0.4}$NiSn$_{0.995}$Sb$_{0.005}$. Inset: (220)-reflection profile for samples produced by spark plasma sintering (SPS) at 1350 (blue), 1250 (red), and 1100 °C (green), and as cast ingot annealed at 700 °C for 10 days (orange). (**b**) Full width at half maximum (FWHM) of six strong and medium reflections (size of symbols represents uncertainty). Peak-broadening $\beta_t$ is due to contributions from average particle size (*D*), strain ε, and instrumental broadening $\beta_o$. Since $D \gg \lambda$, the first term can be neglected. (**c**) Strain vs. processing temperature for SPSed and conventionally annealed samples. The data point for a sample produced by SPS at 1000 °C (brown) is also included. X-ray results for this sample are not shown in (a) and (b) to reduce overlapping data.

The microstructure of the samples fabricated by SPS was shown to consist of nanograins of 50–100 nm in size. The nanograins were produced during recrystallization, whereby the single-phase sample was formed from the mixed-phase sample. After annealing the nanograined sample for 7 days at 700 °C, grain size increased to several microns. On the other hand, there was no additional change in the lattice strain, indicating that the latter is independent of the grain size but depends only on the processing temperature [11].

The influence of strain on the thermoelectric properties is notable. This is manifested in the significant increase in the previously defined power-factor temperature product, *PF*x*T*, as synthesis temperature increased, by as much as 50% across the five samples shown in Figure 4. Among the samples produced by SPS, this resulted in a 20% increase in *zT*, from 1 to 1.2. The large increase in power factor, which is partly responsible for the increase in *zT*, can be attributed to the decrease in charge-carrier density and increase in carrier mobility. This finding is consistent with the increase in size of the bandgap inferred from resistivity measurement [11]. The reduced lattice strain has apparently resulted in the disappearance of most of the in-gap defect states associated with antisite disorder [30] that scatter charge carriers.

Summarizing, the thermoelectric properties and *zT* of n-type (Hf,Zr)NiSn half-Heusler compounds were significantly improved by annealing the materials near the melting point, resulting in the mitigation of lattice strain. The same approach may also improve the performance of other thermoelectric materials.

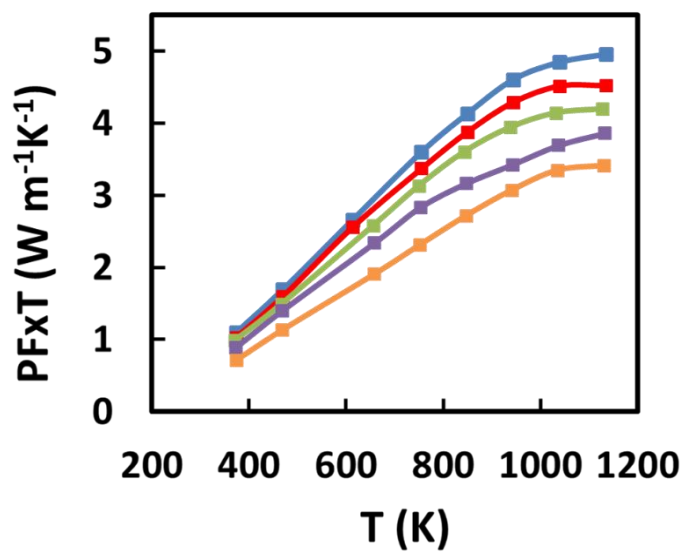

**Figure 4.** *PF*x*T* product as a function of temperature for n-type Hf$_{0.6}$Zr$_{0.4}$NiSn$_{0.995}$Sb$_{0.005}$ HH compounds produced by SPS at 1350 (blue), 1250 (red), and 1100 °C (green); as cast ingots annealed at 1000 °C for two days (purple), and 700 °C for eight days (orange). (Extracted from Reference [11]).

*2.4. Formation of Resonant Dopant States to Enhance zT*

Hybridization between localized dopant states and extended host band states led to band anticrossing in electronegativity-mismatched semiconductor, producing resonant states in some cases [31,32]. Resonant states form a localized sub-band in the density of states (DoS) near the Fermi level, enhancing the DoS. This can result in an increase in electrical resistivity and the Seebeck coefficient, leading to an increase in *zT* in some semiconductors [33–37]. Simonson et al. [37] showed V doping on the Hf/Zr metal site of the half-Heusler compound Hf$_{0.75}$Zr$_{0.25}$NiSn could enhance the Seebeck coefficient. The correlation between the enhanced Seebeck coefficient and increased DoS near the Fermi level was confirmed by specific heat measurement. Stimulated by the success in improving *zT* by reducing the lattice strain, we also explored whether further improvement in *zT* could be obtained by doping one of the *zT* > 1 HH compounds, namely, n-type Hf$_{0.6}$Zr$_{0.4}$NiSn$_{0.995}$Sb$_{0.005}$, with VA group elements V, Nb, and Ta [13].

A systematic investigation of V, Nb, and Ta as resonant dopants was performed on (Hf$_{0.6}$Zr$_{0.4}$)$_{1-x}$M$_x$NiSn$_{0.995}$Sb$_{0.005}$ (M = V, Nb, Ta), where *x* = 0.002, 0.005, and 0.01, respectively. The TE properties of these doped samples are shown in Figure 5. Summarizing, V dopant resonant states were found to increase both the electrical resistivity and the Seebeck coefficient relative to those of undoped compounds. On the other hand, thermal conductivity decreased largely due to increased electrical resistivity. In contrast, Nb and Ta acted as normal (nonresonant) dopants, as evidenced by the decrease in resistivity and the Seebeck coefficient. Overall, the power factor of V-doped compounds remained the same as the undoped compounds, while that of Nb- and Ta-doped compounds decreased. A high *zT* of 1.3 was obtained near 850 K for the (Hf$_{0.6}$Zr$_{0.4}$)$_{0.99}$V$_{0.01}$NiSn$_{0.995}$Sb$_{0.005}$ compounds. The increased *zT* in V doped samples concomitant with the reduced charge—carrier density and increased effective band mass underscores a band-structure mechanism of *zT* enhancement.

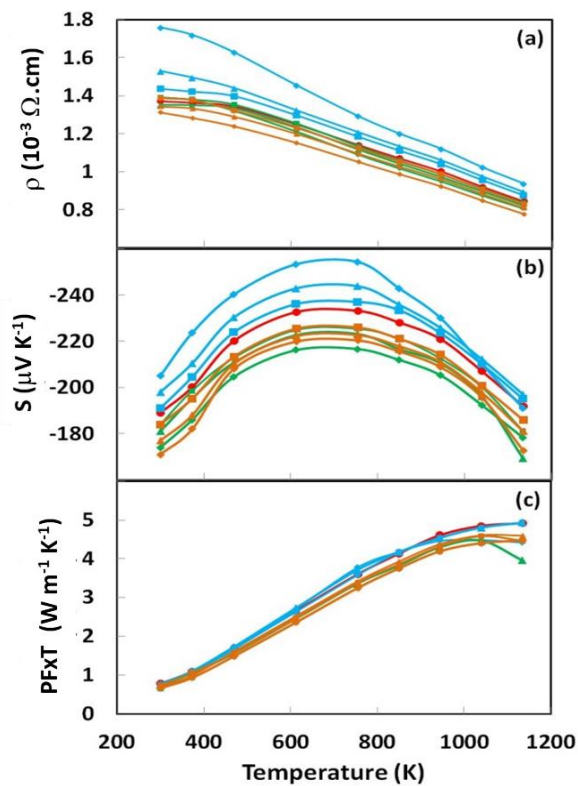

**Figure 5.** (**a**) Electrical resistivity, (**b**) Seebeck coefficient, and (**c**) power factor temperature product of (Hf$_{0.6}$Zr$_{0.4}$)$_{1-x}$M$_x$NiSn$_{0.995}$Sb$_{0.005}$ (M = V, Nb, Ta). $x$ = 0 (red circle); V: $x$ = 0.002 (blue square), 0.005 (blue triangle), 0.01 (blue rhombus); Nb: $x$ = 0.002 (green square), 0.005 (green triangle), 0.01 (green rhombus); Ta: $x$ = 0.002 (orange square), 0.005 (orange triangle), 0.01 (orange rhombus). Overlaps occur in the power factor plots (Extracted from Reference [13]).

The "deep traps impurity level" model of Reference [32] can be utilized to understand the results observed for V, Nb, and Ta dopants. Within the framework of this model, dopant resonant states arise from the interaction of dopant orbitals with the atomic orbitals of the host material. Hybridization of the valence electrons between the Hf(Zr) and X (V, Nb, and Ta) atoms via their nearest-neighbor interaction ν can cause bonding-antibonding splitting, as shown in Figure 6. Following Reference [32], the antibonding level lies above the Hf(Zr) atomic level ε$_{Hf}$ by an amount $ν^2/|ε_X−ε_{Hf}|$, where ε$_X$ denotes the atomic level of X. Since the atomic levels of Nb and Ta lie closer to the Hf level than V based on their relative atomic numbers, the antibonding states that result from Hf-(Nb or Ta) hybridization are expected to lie somewhere further away from the conduction-band edge. As a result, the |ε$_X$−ε$_{Hf}$| term increases from Ta to Nb to V. On the other hand, coupling matrix element *ν*, which depends on the size of the dopant orbital, decreases from Ta to Nb to V. The weaker hybridization between V and Hf also implies a more 'localized' nature of the V-induced hybridized states near the conduction band edge, giving rise to the resonant states that increase electrical resistivity and enhance thermopower.

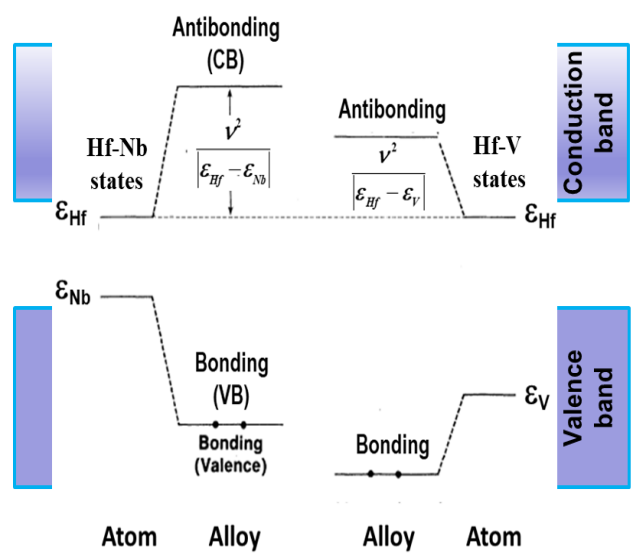

**Figure 6.** Schematic two-state model to illustrate the difference between the positions of the bonding and antibonding levels relative to conduction band (**CB**) and valence band (**VB**) in V- and Nb-doped HfNiSn compounds. See text for more details.

## 3. Conclusion

In recent years, researchers have identified pathways to improve the thermoelectric properties of half-Heusler compounds. The approaches to achieving higher *zT* in half-Heuslers comparable to that of state-of-the-art materials have targeted structural order, microstructure, heavy hole band, band degeneracy, and dopant resonant states. Competitive thermoelectric conversion efficiency and power density are achieved in some TE modules built on half-Heusler compounds. Despite the advantages of half-Heusler compounds in their thermal stability and scalability, important issues such as material lifetime and resilience still need to be overcome prior to their implementation in practical devices.

Advances made in the past few years demonstrated that the thermoelectric properties of half-Heusler compounds can be significantly improved by tuning the compositions of existing compounds. It is amply evident now that there is plenty of room for microstructure design and band-structure engineering in achieving significant improvement in this class of materials. Notably, recent studies have brought electron orbital and valence effects to the fore. Given the vast compositional space of half-Heusler compounds, new compositions may be designed to optimize TE properties as well as validate new ideas. In recent years, material discovery has been well served by the development of various high-throughput methods that employ first-principle thermodynamics, machine learning, ab initio calculations, and combinations of these methods [38–40]. The predictive capability of these methods will continue to improve as the database expands and new algorithms are developed. Research on half-Heusler materials can definitely take advantage of this development. Thus, one can expect a higher level of achievement in *zT* and, equally importantly, the realization of practical thermoelectric generators based on half-Heusler compounds.

**Acknowledgments:** The author thanks Terry Tritt and members of his group of Clemson University for their collaboration in the study of thermoelectric half-Heusler compounds.

**Conflicts of Interest:** The author declares no conflict of interest.